\documentclass[12pt]{article} 
\usepackage[dvips] {epsfig}

\begin{document} 
\setlength{\textwidth}{15 cm} 
\setlength{\textheight}{23cm}   
\baselineskip=21pt 
\parskip=0.5cm 
\parindent=0.2cm
\baselineskip=16pt 
\bibliographystyle{unsrt}
\baselineskip=28pt 

\begin{center}
{\Large{\bf Energy Transfer Efficiency Distributions
in Polymers in Solution During Folding and Unfolding }} \\ 
\vspace{0.2cm}
{\large{\bf Goundla Srinivas  
and Biman Bagchi\footnote[1]
{For correspondence: bbagchi@sscu.iisc.ernet.in}}}\\
Solid State and Structural Chemistry Unit,\\
Indian Institute of Science,\\
Bangalore, India 560 012.\\
\end {center}
\begin {center}
{\large \bf Abstract}
\end {center}
{\it Distribution of fluorescence resonance energy
transfer (FRET) efficiency between the two ends of a Lennard-Jones polymer 
chain both at equilibrium and during folding and unfolding has been 
calculated, for the first time, by Brownian dynamics simulations.  
The distribution of FRET efficiency becomes 
{\bf bimodal} during folding of the extended state subsequent to a temperature
quench, with the width of the distribution for the extended state broader 
than that for the 
folded state. The reverse process of unfolding subsequent to a 
upward temperature jump
shows different characteristics. The distributions show significant viscosity
dependence which can be tested against experiments.}


\newpage

\section {Introduction}

Dynamics of polymer folding and unfolding in solution is a problem of much 
current interest. \cite {kar1,yos,deh} Although some aspects of polymer
folding appears to be understood, detailed experimental
study of the folding scenario (especially
the initial part) is still not available. Given the complexity of the 
problem, the computer simulations could consider only relatively simple 
models, such as a necklace of Lennard-Jones or square-well beads. 
In addition to its own
intrinsic interest, the collapse of 
polymers from poor solvents has served as a theoretical model of  
protein folding in the  
early stages.\cite{dil,pgw1,pgw2,onu,dth,pgw3,boc,dth2,gut}

 In a notable recent development, fluorescence resonance energy 
transfer (FRET) has been combined with single molecule spectroscopic (SMS) 
technique to provide a powerful novel approach to study the dynamics of 
folding. Deniz {\it et al,} \cite {sch2} reported
studies of dynamics of protein folding by observing 
FRET in time domain from a {\em single} donor-acceptor (D-A) pair.
In the measurement of Deniz {\it et al}, the {\em subpopulations} 
\cite {sch2,sch1,wei,hos,bal} of the folded and
denatured states of the protein chymotrypsin inhibitor 2 (CI2) were obtained as 
the concentration of the denaturant (guanidinium chloride) was varied. 
The most interesting result was that at the intermediate
concentration of the denaturant, the distribution 
of Forster efficiency becomes {\it bimodal}. It was concluded
that this bimodal distribution is the signature of the
"two state" nature of  CI2 folding transition. 
We were curious to know whether such a bimodal distribution can be 
observed (a) for simple models like collapsing homopolymers  and
(b) whether it can originate from dynamics alone. To answer these
questions, we have carried
out extensive Brownian dynamics (BD) simulations of FRET in single 
macromolecules. We find,
surprisingly, that even such a simple system shows a bimodal distribution
in the excitation transfer efficiency, for chosen sets of
parameter values. The double peak is similar to the
ones observed in experiments.\cite{sch2} The bimodal distribution is observed
not only during folding but also at equilibrium near the $\theta$ temperature.
The results obtained here should help in designing future experiments.

The usually assumed  mechanism for FRET is the Forster energy 
transfer (FET).\cite {for,sty} 
The rate of this transfer depends on the separation $ (R)$ between the 
energy donor and the energy acceptor. This rate, $k(R)$, can be 
written as\cite{for},
\begin {equation}  
k( R) = k_F \left( {R_F \over  R} \right )^6
\end{equation}  
\noindent where $R_F$ is the Forster radius 
and $k_F$ is the radiative rate, which is typically
in the range $10^{8}$ to $10^{9}$ sec$^{-1}$ for the
commonly  used chromophores. The
rate of energy transfer becomes equal to $K_F$ when 
 $R = R_F$.  For a given D$-$A pair, 
$R_F$ can be obtained by the usual method of overlap between the 
fluorescence and the absorption spectra 
of the D-A pair.\cite {pech}
For commonly used chromophores,
$R_F$ is fairly large, often as large as $50 A$.
This means that the rate is very large when
the donor-acceptor pair is separated by a short
distance. This may actually be a limitation of the
Forster expression which is strictly valid when
the separation between the $D-A$ pair is much
larger than their size.  However, the above
limitation shall have a minor effect in the present study
which is qualitative in nature and aims at exploring
the general aspects of the energy transfer efficiency 
distribution during folding.

In writing 
Eq.(1), the standard averaging over the
orientations of the transition dipole moments has been carried out.
In standard FRET experiments, the macromolecule is doped with a D-A
pair in suitable locations along the chain.\cite{sty}
Excitation transfer can be monitored by following the fluorescence
either from the donor or the acceptor or from both. The
time, $\tau_{rxn}$, taken for the excitation transfer to occur depends 
strongly on the D-A separation $R$, as given by $k(R)$ in Eq.(1). 
For a polymer (or a protein) in solution, both at equilibrium
and during folding/unfolding,
$R$ is not only a fluctuating, 
stochastic function of time, but also varies in a definite
way. In such cases, the distribution of the energy transfer efficiency 
contains non-trivial and  useful information. Note that simulating FRET 
in real proteins is an exceedingly difficult problem -- it is not trivial 
even for a simple homopolymer -- a problem that has eluded theoretical
description even today.

 We define the  FRET efficiency ($\Phi_{F}$) by the following relation,
\begin{equation}
\Phi_{F} =\frac{k(R)}{k(R) + k_{rn}}, 
\end{equation}
\noindent where $k_{rn}$ is the rate which includes the radiative and 
non-radiative
(other than Forster migration) decay rates of the
donor-acceptor pair.
We next define the FRET efficiency distribution $P({\Phi _{F}})$
by the following expression
\begin {equation}  
P({\Phi _{F}}) = {1 \over {\cal N}} \sum_{i=1}^{\cal N} \delta 
(\Phi_F - \Phi_F(\tau_{rxn})) .
\end {equation}  
  The above equation is to be understood in the following fashion. After
choosing a D-A pair at time $ t=0$, the pair is followed till 
the trajectory gets  terminated due to the reaction. We define this
time by $\tau_{rxn}$.
At this time, the existing end-to-end distance ($R$) is used in Eq. (2) to 
calculate
$\Phi_F$. This was followed for ${\cal N}$ independent trajectories for pairs 
chosen from an equilibrium distribution. At the end we distribute the
FRET efficiencies in to the bins of width 0.1. In this way, a continuous 
probability distribution 
$(P({\Phi _{F}}))$ can be obtained from Eq. (3), by taking the 
${\cal N} {\rightarrow \infty} $ limit - in our case, we get a histogram
(figures 1-4).
Similarly, we define the probability distribution of reaction times,
\begin {equation}  
P({\tau _{rxn}}) = {1 \over {\cal N}} \sum_{i=1}^{\cal N} \delta 
(t - \tau_{rxn}) 
\end {equation}

 In this paper, we present the calculations of the distribution of 
$P({\Phi_{F}})$ and $P({\tau _{rxn}})$ for Forster migration among polymer ends 
both at equilibrium and {\em during folding and unfolding}. 
In next section we describe the model and the simulation details. In 
section III we present the results. We close the paper with a few
conclusions in section IV.

\section {Simulation details}

\subsection {The model}

The model polymer chain is made of  connected Lennard-Jones (LJ) beads. 
While this model
homopolymer does not represent complex richness of a protein, it is known
to show interesting folding kinetics.\cite{yos,deh} 
The total potential energy   
can be written as,\cite {yos}
\begin{equation}
U= \sum_{i=2}^{N} \sum_{j=1}^{i-1} u_{LJ}({\bf r}_{ij}) + 
\sum_{i=2}^N u_b (\mid {\bf r}_i-{\bf r}_{i-1} \mid)
\end{equation}
\noindent where $N$ is the number of beads, ${\bf r}_i$ is the position of 
bead $i$, $ r_{ij}=\mid {\bf r}_i - {\bf r}_j \mid, u_{LJ}(r)$ is the 
LJ potential, 
\begin{equation}
u_{LJ}(r) = \epsilon \left[ \left({\sigma \over r} \right)^{12} -
\left({\sigma \over r}\right)^6 \right ] \label {ljeq} 
\end{equation}
\noindent $u_b$ is the bonding potential,
\begin{equation}
u_b =  {3 \kappa k_B  T \over 2 b^2} \sum (\vert {\bf r}_i - {\bf r}_j
\vert - b)^2
\end{equation}
\noindent where $\sigma$ and $\epsilon$ are the LJ collision-diameter and
the well depth, respectively, $k_BT$ is the 
thermal energy and $\kappa$ represents the stiffness of the spring. 
Here, we use $\kappa=9, N = 80$, and $b=\sigma$.
For convenience, we define 
$\epsilon^*=\epsilon/k_BT$. The unit of time, $\tau$, is $b^2/D_0$. 
Thus, $\tilde k_F$ ($\equiv k_F b^2/D_0$) is also dimensionless. 
In this study we have chosen 
$\tilde k_F=1$ and 10. This choice of $\tilde k_F$ corresponds to the
experimentally observed $k_F$ values, a bit
biased towards higher values. For example,
in a solvent with viscosity ($\eta$) equal 1 cp, the radius
of monomer molecules $(R)$ equal to  $4 A^0$, $\tilde k_F=1$ corresponds to a $k_F$ of
$0.76$ ns while $\tilde k_F=10$ corresponds to that of 7.6 ns.\cite {tur} $R$
is scaled by $b$, the bead diameter, as usual. In viscous solvents, the viscosity
can be much higher, and the   $\tilde k_F$ can be even larger than 10.

The time evaluation 
of the polymer chain 
is done according to the following equation of motion, 
\begin{equation}
{\bf r}_j(t + \Delta t) = {\bf r}_j(t) + F_j(t) \Delta t + \Delta X^G(t) 
\end {equation}  
\noindent where ${\bf r}_j(t)$ is the position 
of $j-$th bead at time $t$.
and the systematic force on $j$ is denoted by $F_j(t)$. The random 
Brownian displacement, $\Delta X^G(t)$, is taken from a Gaussian 
distribution with zero  mean and $2 \Delta t$ variance. The 
time step, $\Delta t$, is varied from 0.0001$\tau$ to 0.0005$\tau$.
The scheme of Noguchi and Yoshikawa\cite{yos} has been used to 
investigate the polymer folding.
In this method, the equilibrium configuration 
obtained at $\epsilon^{\star} =0.1$ is quenched by decreasing the temperature 
instantaneously to different values of $\epsilon^{\star}$, higher than $0.62$
which is the $\theta$ temperature in this model. 
The time dependent total energy, the root mean square 
end-to-end distance ($R^2$) and the radius of gyration ($R_g^2$) were all 
monitored to follow the progress of folding.
The results presented here are the average over 10,000
of such trajectories with different initial configuration. More details 
on the simulation scheme can be found elsewhere.\cite {srij}
To simulate FRET , we have 
probed many combinations of $R_F$ and $\tilde k_F$. 
We have selected $R_F=7$, which is near the maximum in $R^2 P(R)$ (we denote 
it by $R_0$), where $P(R)$ is the end-to-end distribution.  
Another important parameter which affects FRET is $\tilde k_F$. 
Large $\tilde k_F$ values result in the higher efficiency of FRET. In 
this study we have mostly dealt with  $\tilde k_F$=10.
 
\subsection {Time scales}

 FRET in polymers involves {\em several} different
time scales.  Two time scales, $\tau_{Ffold}$ and $\tau_{Fun}$,
are required to describe the average survival probability of FRET
in the folded and unfolded states,
respectively.\cite {srij} 
These two time scales are widely separated from each 
other, due to the sensitivity of survival time to the separation
between the two ends. 
The third relevant time in this problem is $\tau_{qfold}$, the time required 
for the polymer to fold subsequent to the quench. 
For FRET to be useful in the study of
folding, this $\tau_{qfold}$ should be intermediate and well-separated
from the other two times. 
Two additional time scales $k_F^{-1}$ and $b^2/D_0$ come from Forster energy
transfer and Brownian dynamics, respectively. 
While $k_F$ is fixed,
$b^2/D_0$ can be varied by changing viscosity ($\eta$).

\section {Results and Discussion}

  Results for the distribution {\it during} folding process 
(subsequent to the quench from $\epsilon^*=0.1$ to
$\epsilon^*=0.8$ ) are shown 
in { figure 1}. In this figure the simulated probability distribution 
of FRET efficiency $P(\Phi_F)$ during folding is plotted at $R_F=7$ 
and $\tilde k_F=10$. One sees a clear bimodal distribution in the
FRET efficiency. The first peak at low efficiencies arise from the extended
state while the one at high efficiencies arise from the folded
configurations. Note also that the distribution for the extended state is
broad while that from the collapsed state is narrow. This is expected
on physical grounds and has been observed in experiments. This bimodality
is found to depend critically on the value of $k_{rn}$ which 
is a consequence of several competing time scales in the FRET.
	
Results presented in figure 1 can be better understood from {figure 2}
where we have plotted the {\it equilibrium} FRET efficiency distribution
in the extended (unfolded) and the collapsed (folded) states, at
$R_F=7$ and $\tilde k_F=10$. It is  observed
that the time taken for FRET is much larger in the unfolded
state (Fig. 2a), compared to that in the folded state (Fig. 2b). 
This is reflected in the (ensemble averaged) survival probability
($S_p(t)$) of D-A pair (not shown), which is extremely fast in the 
collapsed state and very slow in the extended state.
The reason for the observed (nearly) exponential distribution for
the unfolded state (Fig. 2a) lies in the choice of $R_F$. Since  
maximum probability for end-to-end distance  (obtained by maximizing $R^2P(R)$)
is at $R\approx 7.3$, for the given temperature and interaction strength. Thus,
at $R_F=7$, there is significant population already at this distance.
If we change $R_F$ to small (like $R_F=1$) or large (like $R_F=10$)
values, the exponential distribution will be replaced by a Gaussian type
distribution. This sensitivity of the distribution to $R_F$ can be exploited
in experiments.
The distribution of reaction efficiencies in the folded state (Fig. 2b), 
however, is {\em not} exponential. The initial fast fall in the probability is 
followed by a somewhat slower decay.

Both at high ($\epsilon^*\approx 0.1-0.3$) and low temperatures
($\epsilon^*\approx 0.8-0.9$) the equilibrium
$P(\Phi_F)$ shows a single peak at lower and higher 
FRET efficiencies, respectively. 
However, at intermediate temperatures($\epsilon^* \approx 0.5-0.6$), the
equilibrium FRET efficiency distribution $(P(\Phi_F))$ again shows a bimodal 
distribution. This is shown in { figure 3}, where 
the equilibrium $P(\Phi_F)$
is plotted at $\epsilon^*=0.5$. The emergence of bimodality 
in $P(\Phi_F)$, even at equilibrium, 
is due to the closeness to the $\theta$ temperature 
(${\epsilon_{\theta}^*=0.62}$). 
This indicates that a "two-state" model exists in this simple system of 
homopolymer chain, near the $\theta$ temperature . 

 The above observations seem to provide the following interpretation of
 the observed bimodality in figure 1. As the polymer collapses subsequent to 
 the quench, the polymer passes through a succession of configurations. The
 initial configurations correspond to the extended state. The polymer passes 
 through the intermediate state rather fast which shows the dearth of population
 at intermediate $\Phi_F$, giving rise to the bimodality.

The effect of viscosity ($\eta$) can be studied by varying 
the dimensionless rate
$\tilde k_F$, defined as $k_Fb^2/D_0$.
At constant
$k_F$, large values of $\tilde k_F$ represent solution of high viscosity 
and vice versa.
In { figure 4}, $P(\Phi_F)$ is plotted against the FRET efficiency at two
very different values of $\tilde k_F$. Open bars show the result for 
$\tilde k_F=1$, while the filled bars represent that for the 
$\tilde k_F=10$. Figure 4 shows  that viscosity can have a dramatic
effect on FRET efficiency distribution.

 During the unfolding (when the temperature instantaneously  raised to
$\epsilon^*=0.1$ from $\epsilon^*=0.8$ ),
$P(\Phi_F)$ shows a large peak at higher efficiencies which is accompanied 
by a tail (of smaller peaks) towards lower efficiencies. 
The probability distribution of reaction times ( subsequent to a
temperature quench)
$P(\tau_{rxn})$ also shows a bimodal distribution.
We have found that one  can efficiently study the dynamics of the 
initial stages of 
folding by placing the pair near the chain end where the nucleation
of the collapse starts in nearly all the cases that we studied.
In this case distribution of the reaction efficiency is not as strongly
bimodal as in the case previously mentioned. 

We have monitored the variation in the mean square radius
and the energy during the polymer folding as a function of time.
They show somewhat different behavior.  $\langle R^2 \rangle$ and 
$\langle R_g^2 \rangle$ starts decaying  only after an initial 
characteristic time delay. The total energy of polymer
chain starts decaying immediately after the quench and 
continues to do so till it reaches the final stable minimum energy
configuration. 
This is because initial decrease of energy does not
require change in $R_{0}$ -- it occurs by establishing favorable contacts.

\section {Conclusion}

To conclude, we have shown in this work that a bimodal
distribution of  excitation transfer efficiency and of reaction times emerge
during folding and unfolding of model homopolymers in solution. The
distribution looks surprisingly similar to the ones observed recently in the
folding of real proteins by single molecule spectroscopy. The extent 
of bimodality is found to depend on the values
of the Forster parameters ($k_F$ and $R_F$),$ k_{rn}$
and the value of the diffusion coefficient (that is, the viscosity).
Thus, a study like the one performed here can be useful in designing
FRET experiments via single molecule spectroscopy.
The present study suggests several exciting future problems. Both for
folding and unfolding, one can initiate the FRET process after a suitable
time lag ($\tau$) following the quench and can thus 
obtain a {\it two dimensional
distribution} of $P(\Phi_F,\tau)$, like in NMR or ESR.
FRET may also be able to differentiate between different
collapsed states, like rod and toroid.\cite{yos,deh} Simulations
of this can be carried out by using a stiff polymer chain.\cite{yos,deh}
Simulation study by using more realistic models and also by
incorporating the solvent molecules implicitly, may reveal more information. 
Further work in these directions is under progress.  Finally, simulations
of $P(\Phi_F)$ during folding of real proteins remains a challenging task.

{\Large \bf Acknowledgments}
 
It is pleasure to thank Prof. A. Yethiraj for the help and discussion.
The financial support from CSIR, New Delhi, India and DST India is 
gratefully acknowledged. G. Srinivas thanks CSIR for a research 
fellowship.

\newpage
\noindent {\large \bf Figure captions:}

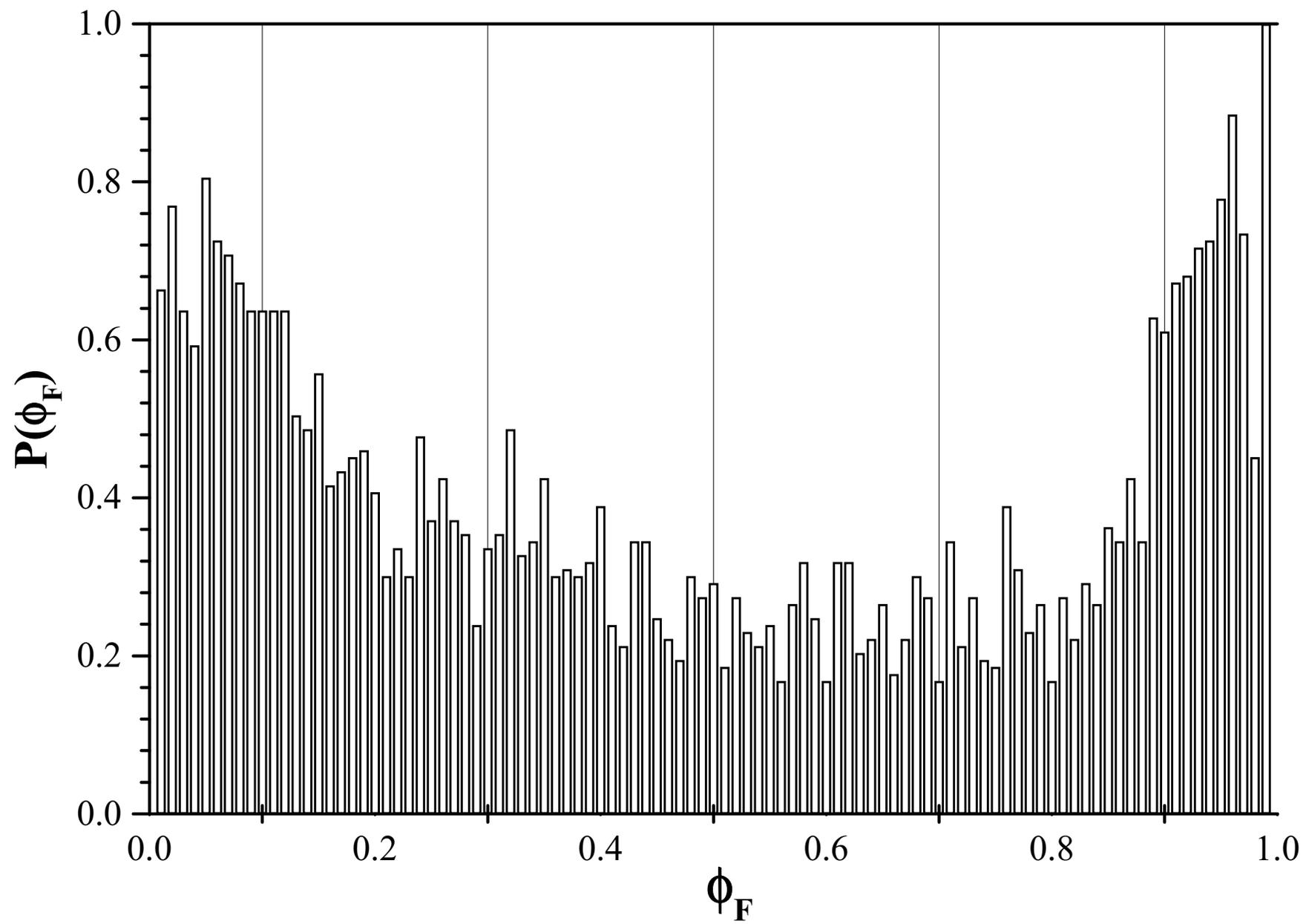
\noindent{\bf Figure 1.} 
The distribution of FRET efficiency ($P(\Phi_F$)), 
subsequent to a temperature 
quench (from $\epsilon^*=0.1$ to  $\epsilon^*=0.8$ ), obtained from BD 
simulations is shown at  $R_F=$ 7,  $\tilde k_F=$10 and N=80. 
This demonstrates the emergence of bimodality in $P(\Phi_F)$.
Here $k_{rn}$ is fixed as 0.0001; the bimodality
is present at other values also, but sharpness depends on the
magnitude of the radiative rate.

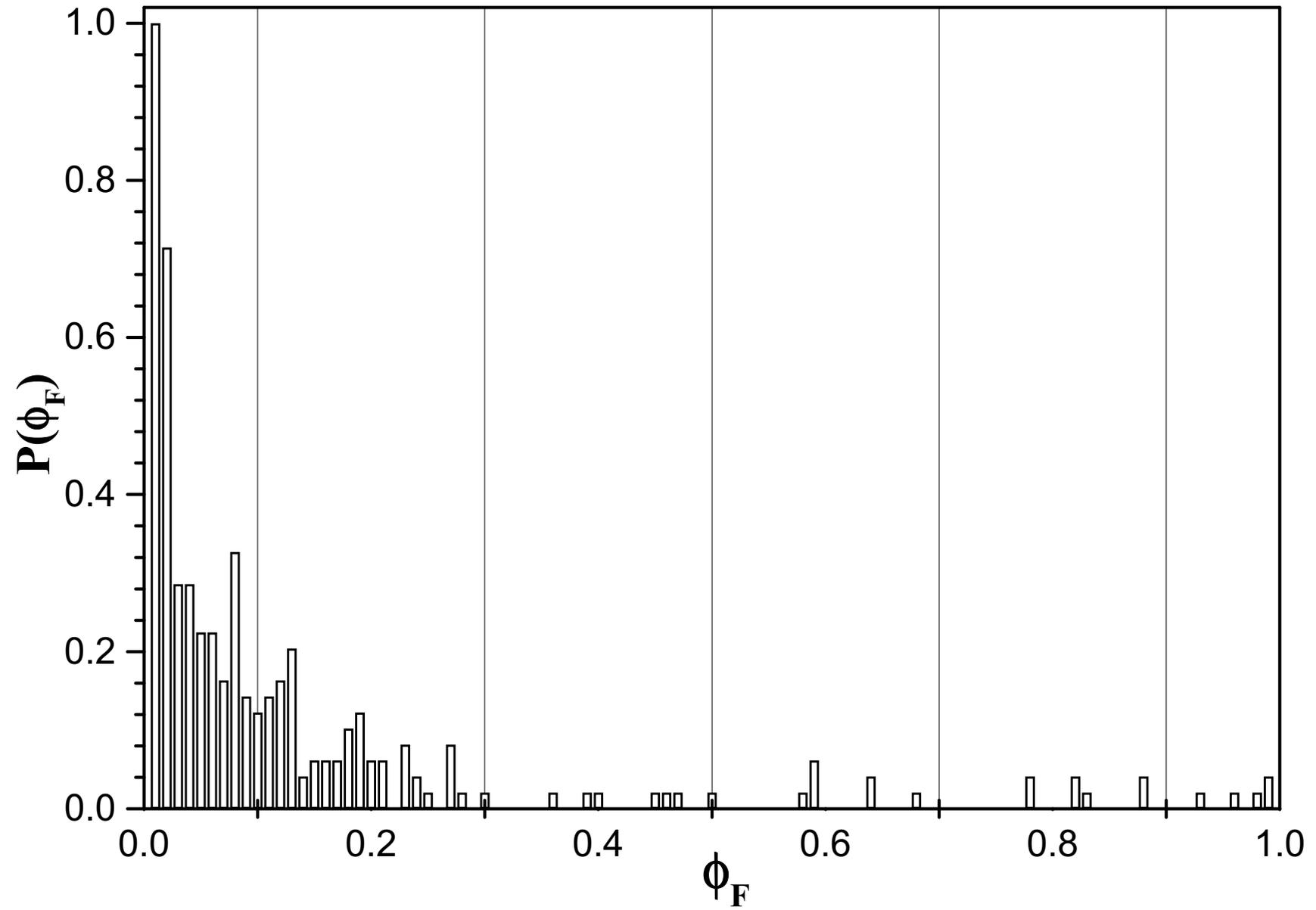
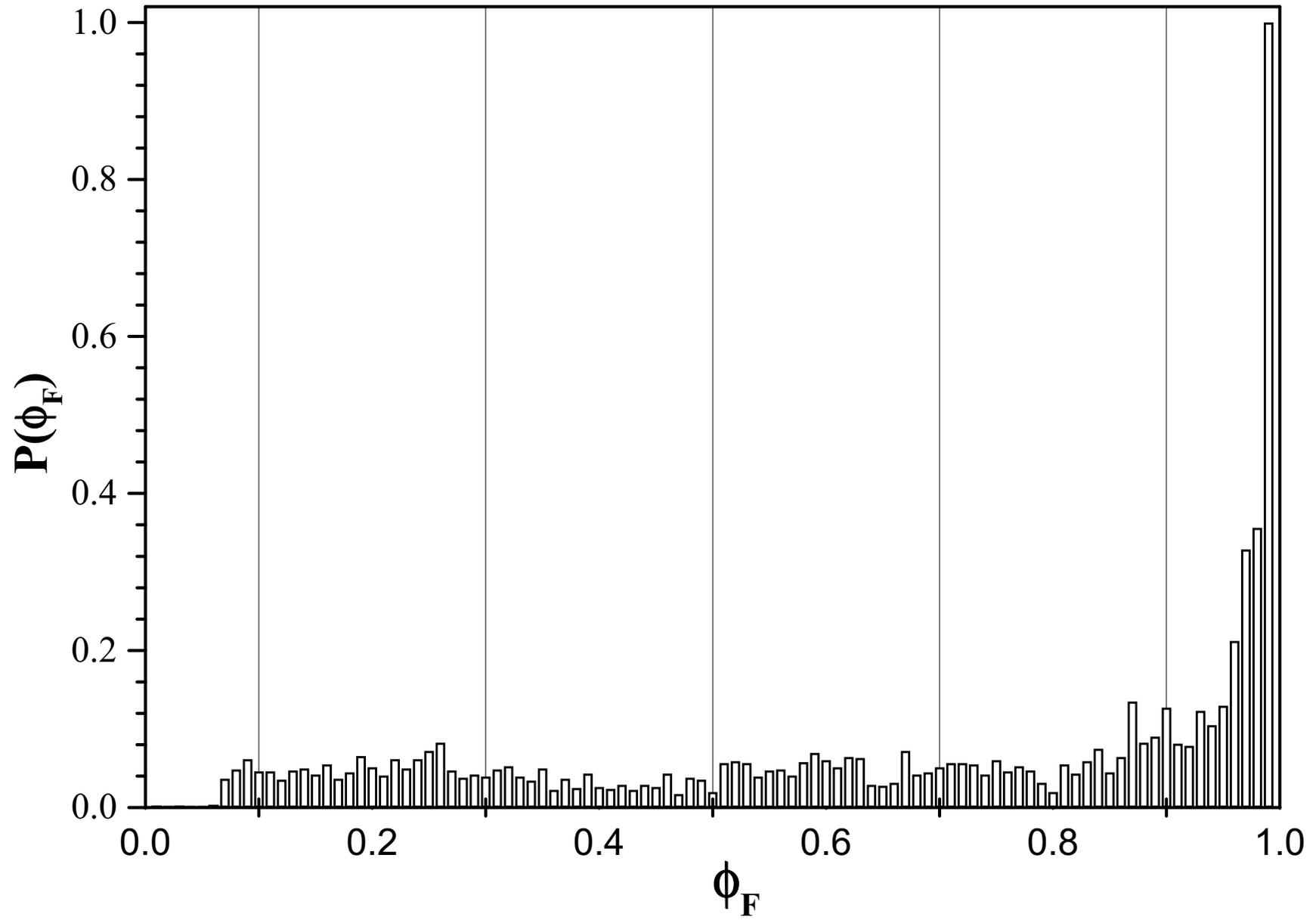
\noindent{\bf Figure 2.} The equilibrium FRET efficiency ($\Phi_F$) 
distribution 
is shown for (a) $\epsilon^*=0.3$ and (b) $\epsilon^*=0.8$. $\tilde k_F=10$ and 
$R_F=7$ while  $k_{rn}$ is fixed as 0.01.  
 
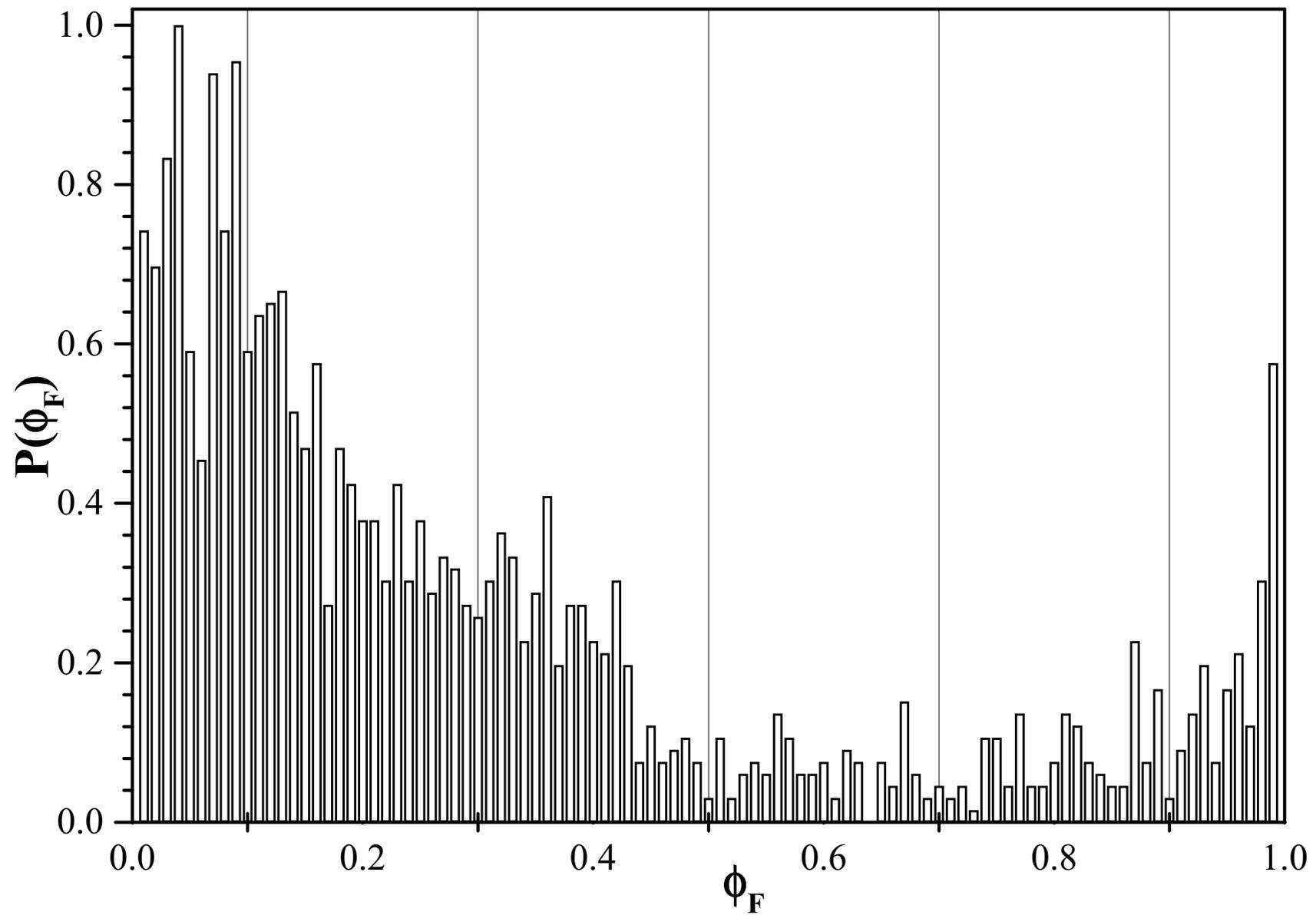
\noindent{\bf Figure 3.} The equilibrium FRET efficiency 
distribution ($P(\Phi_F$)) is shown for $\epsilon^*=0.5$, which is 
close to the $\theta$ temperature ($\epsilon_{\theta}^*=0.62$). 
$\tilde k_F$, $R_F$ and $k_{rn}$ are same as in figure 2.  
This figure demonstrates the emergence  of bimodality even at equilibrium.

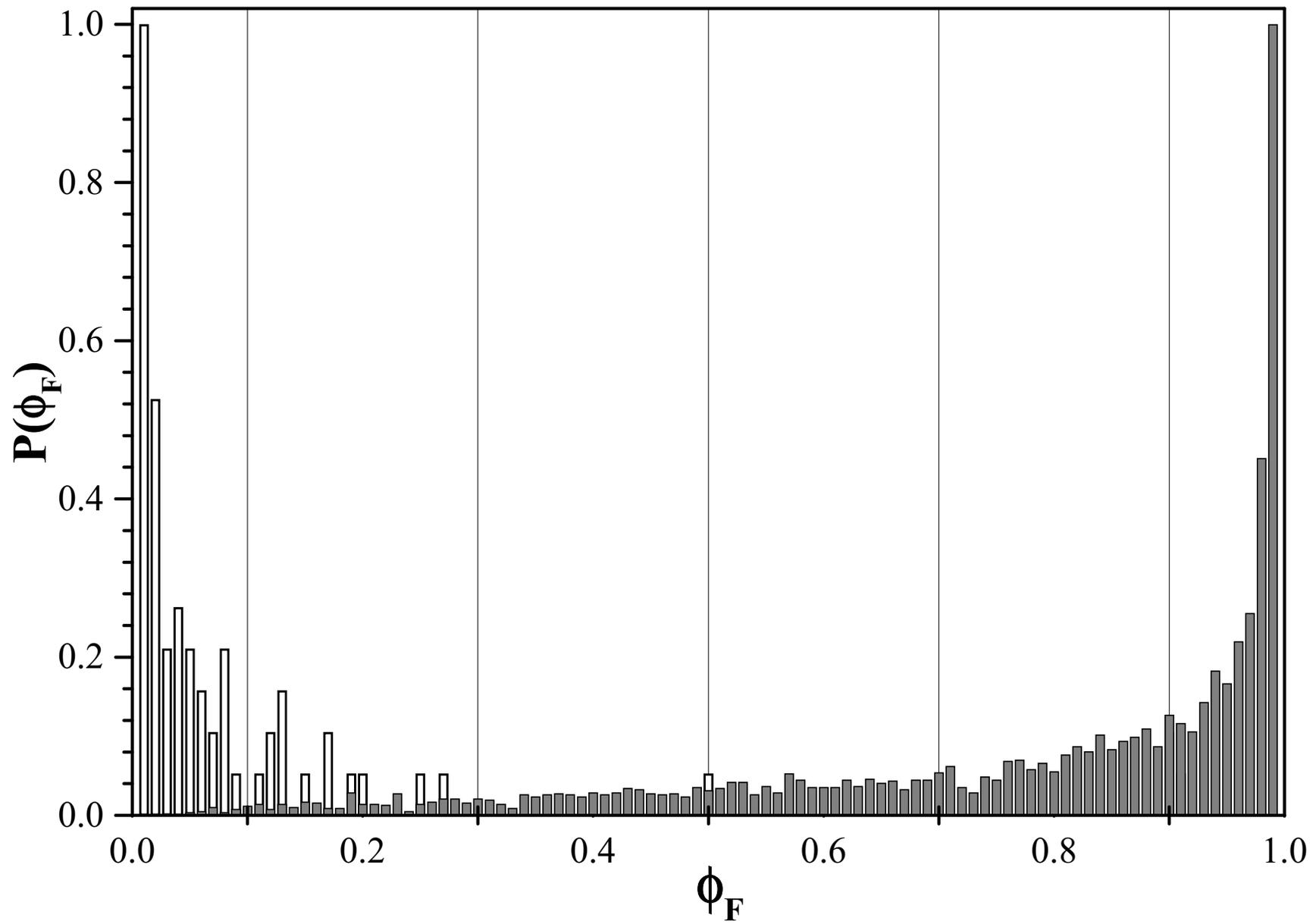
\noindent{\bf Figure 4.} FRET efficiency 
distribution ($P(\Phi_F$)), subsequent to quench ( from $\epsilon^*=0.1$ to 
$\epsilon^*=0.8$) is shown for $\tilde k_F=$1 (open bars) and 
$\tilde k_F=10$ (filled bars). 
$R_F=7$ and $k_{rn}$ are same as in figure 2.  

\end {document}